\documentclass{elsart}

\usepackage{amssymb}
\usepackage[all]{xy}
\begin{document}
\begin{frontmatter}

\title{Collective dipole oscillations in hydrogenated metals}
\author{J.S.Brown}
\address{Clarendon Laboratory, University of Oxford, Parks Rd. OX1 3PU, UK}

\begin{abstract}
It is observed that low-lying transitions of an interstitial hydrogen adatom on a metallic lattice correspond classically to dipoles oscillating at frequencies where band electrons typically have a low electromagnetic absorption. Such transitions hence give rise to an essentially unscreened time-dependent perturbation of the lattice potential experienced by neighbouring hydrogen states. It is shown that above a certain hydrogen concentration, coherently phased oscillations at adjacent sites may result in a dipole-dipole interaction larger than the excitation energy. This would give rise to a new ground state for the ensemble of coupled oscillators and a reduction in lattice symmetry below some phase transition temperature. An explicit calculation is carried out for PdH and PdD, and indirect evidence for the existence of the postulated phase is presented.
\end{abstract}

\end{frontmatter}

\section{Introduction}
\label{intro}
It is generally accepted that certain properties of substoichiometric transition metal hydrides are not fully explained by existing models. In addition to the long-standing reverse isotope dependency of some superconducting transition temperatures, it is not clear whether the high mobility of protons in certain metals is fully accounted for by phonon processes. Furthermore, neutron scattering studies \cite{Kar03} appear to suggest a quantum phase correlation between hydrogen adatoms at the femtosecond timescale, which is collaborated by EELS studies \cite{Oku01} indicating hydrogen delocalization on Ni(111) surfaces. Finally, and perhaps most intriguingly, beam-foil experiments \cite{Kas02,Rai02} have revealed anomalously large fusion cross-sections when low energy deuterons traverse foils composed of an wide range of metallic elements \cite{Cze04,Huk06}.  
A recent attempt \cite{Kal05} to explain the latter effect in terms of phonon exchange processes failed to account for the full magnitude of the observed anomaly. We were accordingly led to look for a further, hitherto neglected, generic effect whose inclusion in the model Hamiltonian might yield a more complete description of these systems.
\section{Hydrogen adatoms on a periodic lattice}
\label{model}
The many-body Hamiltonian governing the interactions of interstitial hydrogen adatoms amongst themselves and with a periodic lattice of arbitrary symmetry can be expressed as
\begin{equation}
\label{Hall}
\begin{array}{ccccc}
\epsilon_i c^\dagger_{R,i} c_{R,i} & + & M_{i,j,\omega,\Gamma,\Delta} (b^\dagger_{\omega, \Gamma} + b_{\omega, \Gamma}) c^\dagger_{i,R+\Delta} c_{j,R} & + & \omega b_\omega ^\dagger b_\omega
\\
& + & V_{i,j,\Delta} c^\dagger_{R+\Delta,j'} c^\dagger_{R,j} c_{R+\Delta,i'} c_{R,i} & &
\end{array}
\end{equation}
where $c^\dagger_{R,i}$ creates a hydrogen adatom in state $i$ at site $R$ and
$\epsilon_i$ is an eigenvalue of the single-particle Schr\"odinger equation solved in the region of the site. This is invariably a high symmetry (e.g. octahedral, tetragonal) point of the primitive cell at which the total effective lattice potential is a local minimum. 
\\
$\Delta$ denotes a member of the set of near neighbour displacements. $b^\dagger_{\omega,\Gamma}$ creates a phonon of energy $\omega$ and polarization $\Gamma$. The elements of the phonon-coupling matrix $M$ are in general a complicated function of all five indices. The elements of $V$ can be derived, in principle, from the photon propagator kernel in the dielectric medium of the band electrons. Repeated indices indicate an implicit summation over that variable.
\\
The site-diagonal $\Delta = 0$ elements of the hydrogen-phonon coupling $M$ can be transformed into an effective self-trapping effect, whereby each hydrogen creates a static phonon displacement field of short range. The price we pay for the simplification resulting from the neglect of phonon processes is that the site energies $\epsilon_i$ and wavefunctions $\varphi_i(r-R)$ become somewhat dependent on the spatial distribution of occupancy at neighbouring sites. 
\\
The $\Delta \neq 0$ elements of $M$ are responsible for site-hopping. This can be represented by a term of the form 
$t\sum  c^\dagger_{i,R + \Delta} c_{R,i}$ where $t$ is on the order of 1 meV and the sum is taken over nearest neighbour $\Delta$ only.
\\
The site-diagonal (electrostatic repulsion) elements $V_{i,j,0}$ are of the order of a few eVs and effectively prohibit multiple occupancy of sites in all conceivable lattices.
\\
At small energies corresponding to transitions of the band electrons about the Fermi level $\epsilon_F$, the interaction potential $V$ is reasonably well described by Thomas-Fermi screening. With the values of $N(\epsilon_F)$ typical \cite{Cha83} of the metals capable of absorbing large amounts of hydrogen, neighbours are effectively screened from one another and the intersite interaction (i.e. $\Delta \neq 0$) is of the order of a few meV only. However, this holds only to the extent that neighbouring occupants both reside in their respective ground states (i.e. $i(t)=j(t)=0$).  The first excited states in the potential well experienced by hydrogen adatoms are typically \cite{Oku01,Kri94,Sun04} on the order of 0.1 eV above the ground state. The corresponding transition frequency lies in a region where the electromagnetic absorption coefficient K of a metal is typically of the order of a few tens of $cm^{-1}$ and where the attenuation over a distance of a few \AA \ is insignificant. This means that, for $i \neq j$, the intersite interaction is {\it not} subject to the exponential screening factor. Accordingly:
\begin{equation}
\label{Vij}
V_{i,j}(r,r') \approx {e^2 \over |{\bf r}-{\bf r'}|} e^{-4\pi e^2 N(\epsilon_F) |{\bf r}-{\bf r'}|\delta_{ij}}
\end{equation}
where we have replaced the displacement vector index $\Delta$ by the continuous functional dependence over real space.
\\
This crucial distinction between {\it slow} displacements of the hydrogen adatom and {\it rapid} transitions between relatively long-lived widely separated levels appears to have been overlooked in previous models of the hydrogen-metal system. We accordingly adopt the hypothesis that, by taking proper account of the frequency-dependence of the interaction between neighbours in the model Hamiltonian, we may be able to account for at least some of the anomalous effects mentioned in the introduction above. 
\\
Incorporating all phonon effects and static nearest neighbour repulsion into the state energy as in \cite{Kri94}, the original Hamiltonian (\ref{Hall}) becomes:
\begin{equation}
\label{Hsimp}
{\overline \epsilon_i} n_{R,i} + U n_R (n_R-1) + t c^\dagger_{R+\Delta,i} c_{R,i} 
+ V_{i,j \neq i,\Delta \neq 0} c^\dagger_{R+\Delta,j'} c^\dagger_{R,j} c_{R+\Delta,i'} c_{R,i}
\end{equation}
where we have made the replacement $n = c^\dagger c$, the effective on-site repulsion energy is represented by a constant $U$ and $\overline{\epsilon_i}$ are the neighbour-dependent self-trapped energies characterized by a bandwidth on the order of a few meV.
\\
Retaining just the lowest order term in the multipole expansion of the Coulomb operator $1 \over |{\bf r} - {\bf r'}|$, we have the following explicit form for the dominant (dipole-dipole) contribution to the expectation value of simultaneous (i.e. photon mediated) transitions between two pairs of states at sites with position vectors ${\bf R}$ and ${\bf R'} = {\bf R} + {\bf \Delta}$
\begin{equation}
\label{dipdip}
\langle j,j' | V_\Delta |i,i'\rangle = \Delta^{-3} \langle j |{\bf r}|i\rangle \cdot \langle j' |{\bf r'}|i'\rangle
- 3 \Delta^{-5} \langle j |{\bf \Delta.r}|i\rangle \langle j' |{\bf \Delta.r'}|i'\rangle
\end{equation}
This is the first-order correction to the energy of a pair of neighbours due to their oscillatory dipole-dipole interaction. It is clear that a significant second-order effect of the dipole-dipole interaction will be to break the cubic symmetry of the lattice such that $a_\parallel > a_\perp$, since any elongation along the dipole axis will tend to increase the dipole moment of the odd-parity states as well as reduce their (single-particle) energy relative to the other members of their respective multiplets.
\\
The lowest energy configurations of O-site dipoles in an fcc host and T-site dipoles in a bcc host are shown below:
\\
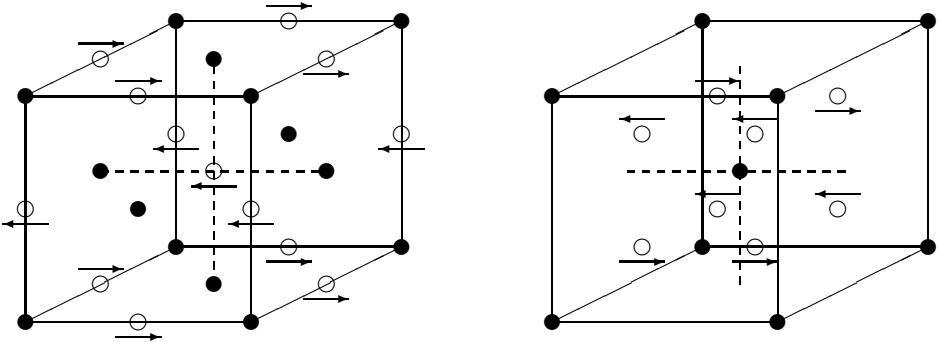
\begin{figure}[!h]
\caption{Dipole alignment in cubic cells}
\setlength{\unitlength}{1mm}
\begin{picture}(160,60)

\put(5,5){\line(1,0){30}}
\put(5,5){\line(0,1){30}}
\put(35,5){\line(0,1){30}}
\put(5,35){\line(1,0){30}}
\put(25,15){\line(1,0){30}}
\put(25,15){\line(0,1){30}}
\put(55,15){\line(0,1){30}}
\put(25,45){\line(1,0){30}}
\put(5,5){\line(2,1){20}}
\put(35,5){\line(2,1){20}}
\put(5,35){\line(2,1){20}}
\put(35,35){\line(2,1){20}}
\put(5,5){\circle*{2}}
\put(5,35){\circle*{2}}
\put(35,5){\circle*{2}}
\put(35,35){\circle*{2}}
\put(20,20){\circle*{2}}
\put(25,15){\circle*{2}}
\put(25,45){\circle*{2}}
\put(55,15){\circle*{2}}
\put(55,45){\circle*{2}}
\put(40,30){\circle*{2}}
\put(30,10){\circle*{2}}
\put(30,40){\circle*{2}}
\put(15,25){\circle*{2}}
\put(45,25){\circle*{2}}
\multiput(15,25)(2,0){15}{\line(1,0){1}}
\multiput(30,10)(0,2){15}{\line(0,1){1}}
\put(30,25){\circle{2}}
\put(33,23){\vector(-1,0){6}}
\put(15,10){\circle{2}}
\put(12,12){\vector(1,0){6}}
\put(45,10){\circle{2}}
\put(42,8){\vector(1,0){6}}
\put(15,40){\circle{2}}
\put(12,42){\vector(1,0){6}}
\put(45,40){\circle{2}}
\put(42,38){\vector(1,0){6}}
\put(5,20){\circle{2}}
\put(8,18){\vector(-1,0){6}}
\put(25,30){\circle{2}}
\put(28,28){\vector(-1,0){6}}
\put(35,20){\circle{2}}
\put(38,18){\vector(-1,0){6}}
\put(55,30){\circle{2}}
\put(58,28){\vector(-1,0){6}}
\put(20,5){\circle{2}}
\put(17,3){\vector(1,0){6}}
\put(20,35){\circle{2}}
\put(17,37){\vector(1,0){6}}
\put(40,15){\circle{2}}
\put(37,13){\vector(1,0){6}}
\put(40,45){\circle{2}}
\put(37,47){\vector(1,0){6}}

\put(75,5){\line(1,0){30}}
\put(75,5){\line(0,1){30}}
\put(105,5){\line(0,1){30}}
\put(75,35){\line(1,0){30}}
\put(95,15){\line(1,0){30}}
\put(95,15){\line(0,1){30}}
\put(125,15){\line(0,1){30}}
\put(95,45){\line(1,0){30}}
\put(75,5){\line(2,1){20}}
\put(105,5){\line(2,1){20}}
\put(75,35){\line(2,1){20}}
\put(105,35){\line(2,1){20}}
\put(100,25){\circle*{2}}
\put(75,5){\circle*{2}}
\put(75,35){\circle*{2}}
\put(105,5){\circle*{2}}
\put(105,35){\circle*{2}}
\put(95,15){\circle*{2}}
\put(95,45){\circle*{2}}
\put(125,15){\circle*{2}}
\put(125,45){\circle*{2}}
\put(87,15){\circle{2}}
\put(84,13){\vector(1,0){6}}
\put(87,30){\circle{2}}
\put(90,32){\vector(-1,0){6}}
\put(102,15){\circle{2}}
\put(99,13){\vector(1,0){6}}
\put(102,30){\circle{2}}
\put(105,32){\vector(-1,0){6}}
\put(97,20){\circle{2}}
\put(100,22){\vector(-1,0){6}}
\put(97,35){\circle{2}}
\put(94,37){\vector(1,0){6}}
\put(113,20){\circle{2}}
\put(116,22){\vector(-1,0){6}}
\put(113,35){\circle{2}}
\put(110,33){\vector(1,0){6}}
\multiput(85,25)(2,0){15}{\line(1,0){1}}
\multiput(100,10)(0,2){15}{\line(0,1){1}}
\end{picture}
\end{figure}
\\
Solid circles denote metal cores and open circles denote hydrogen sites. 
\\
With randomly distributed fractional occupancy $\lambda$ and complete dipole alignment, the sum over all fcc O-sites yields a dipole-dipole energy per hydrogen of
$\approx -4.4e^2 d^2 a_\perp^{-3} \lambda$. The corresponding result for T-sites in a bcc host is
$\approx -21.2e^2 d^2 a^{-3} \lambda$.
\\
\\
Whether, and to what degree such an alignment actually takes place depends upon the temperature and the relative magnitudes of the transition and dipole-dipole energies. We will describe the outlines of an approximate approach to this problem in the following section.
\section{Mean-field theory}
We consider the effective Hamiltonian  for a pair of neighbouring sites $A,B$. In the interests of clarity and brevity we will limit ourselves to just two accessible states. These need not necessarily belong to the two lowest multiplets. The extension to a more realistic number of states is relatively straightforward.
\begin{equation}
\begin{array}{cc}
\mathcal{H}_{eff} =& {\overline \epsilon_0} ( n_{A,0} + n_{B,0} ) + 
{\overline \epsilon_1} ( n_{A,1} + n_{B,1} )
\\
+& \alpha V (c^\dagger_{B,0} c^\dagger_{A,1} c_{B,1} c_{A,0} + c^\dagger_{B,1} c^\dagger_{A,0} c_{B,0} c_{A,1})
\end{array}
\end{equation}
where $V$ is the nearest-neighbour dipole-dipole interaction given by (\ref{dipdip}) and $\alpha$, the off-diagonal order parameter of the model, is the thermodynamic average over the ensemble of the term in $V$ summed over all sites.
\\
Each site can be either unoccupied or contain a hydrogen in state A or B. There are hence nine possible states for the two sites in this model. The chemical potential $\mu$ must be included, since we assume an external reservoir of hydrogen and treat the hydrogen occupancy fraction $\lambda$ as a state variable. The effective two-site energy levels are the eigenvalues of the following matrix:
\begin{equation}
\label{nine}
\begin{array}{c}
\mathcal{H}_{eff} - \mu \sum n_{i,R} =
\\
\left[
\begin{array}{ccccccccc}
0 & 0 & 0 & 0 & 0 & 0 & 0 & 0 & 0
\\ 
0 & \epsilon_0 - \mu & 0 & 0 & 0 & 0 & 0 & 0 & 0
\\ 
0 & 0 & \epsilon_1 -\mu & 0 & 0 & 0 & 0 & 0 & 0
\\ 
0 & 0 & 0 & \epsilon_0 -\mu& 0 & 0 & 0 & 0 & 0
\\ 
0 & 0 & 0 & 0 & 2\epsilon_0 -2\mu& 0 & 0 & 0 & 0
\\ 
0 & 0 & 0 & 0 & 0 & \epsilon_0+\epsilon_1 -2\mu& 0 & \alpha^* V & 0
\\ 
0 & 0 & 0 & 0 & 0 & 0 & \epsilon_1 -\mu& 0 & 0
\\ 
0 & 0 & 0 & 0 & 0 & \alpha V & 0 & \epsilon_1 + \epsilon_0 -2\mu& 0
\\ 
0 & 0 & 0 & 0 & 0 & 0 & 0 & 0 & 2\epsilon_1 - 2\mu
\end{array}
\right]
\end{array}
\end{equation}
By inspection one sees that the two-site partition function is
\begin{equation}
\mathcal{Z}= 1+2 e^{\beta(\mu - \epsilon_0)} + 2 e^{\beta(\mu - \epsilon_1)} + e^{2\beta(\mu - \epsilon_0)}
+ e^{2\beta(\mu - \epsilon_1)} + e^{-\beta \epsilon_+} + + e^{-\beta \epsilon_-}
\end{equation}
where $\beta \equiv (kT)^{-1}$ and 
\begin{equation}
\label{epm}
\epsilon_{\pm} = \epsilon_0 + \epsilon_1 - 2\mu \pm \alpha V
\end{equation}
\begin{equation}
\lambda = 2\left[ e^{\beta(\mu - \epsilon_0)} + e^{\beta(\mu - \epsilon_1)} +  e^{2\beta(\mu - \epsilon_0)}
+ e^{2\beta(\mu - \epsilon_1)} + e^{-\beta \epsilon_+} + + e^{-\beta \epsilon_-} \right ] / \mathcal{Z}
\end{equation}
\begin{equation}
\label{alpha}
\alpha = \Lambda \left[e^{-\beta \epsilon_-} - e^{-\beta \epsilon_+} \right ] / \mathcal{Z}
\end{equation}
where $\Lambda$ is the symmetry-dependent factor resulting from sum over all sites.
\\
The system of equations (\ref{nine}-\ref{alpha}) can be solved iteratively until self-consistency is obtained.
\section{Application to PdH and PdD}
\label{PdH}
These systems have been the subject of a great deal of experimental and theoretical work. Adiabatic Pd-H pair potentials that yield a spectrum of lower excited states agreeing well with empirical determinations of the optical transition frequencies have been obtained from {\it ab initio} DFT calculations \cite{Kri94,Dye05}. We accordingly used the pair potential VII of \cite{Kri94} to solve the Schr\"odinger equation for a single hydrogen adatom over a real space mesh of width of 0.04 \AA. The lowest states were all found to be centred at the octahedral (O-site) symmetry point equidistant from six metal ions. Some higher states are centred on the tetragonal (T-site) symmetry points located at the midpoint between the O-sites and the metal cores, but these lie higher in energy than those we will be considering. The $A_{1g}$ singlet ground state has even parity (+++) with respect to all three axes. The first excited state is a $T_{1u}$ triplet of odd parity with respect to one axis. The second excited state is a $A_{2g}$ triplet of odd parity with respect to two axes. Some of our results are tabulated below:
\\[1cm]
\begin{tabular}
{|c|cccc|cccc|}
\hline
Isotope    				& & proton         	& 	&   & & deuteron     &	&  \\
\hline
Occupancy $\lambda$ 			& 0.25  & 1.0 	& 1.0 	& 1.0 & 0.25 & 1.0&1.0	&1.0\\
$a_y$,$a_z$ (\AA)   			& 3.94  & 4.07	& 4.07	& 4.07& 3.94& 4.07&4.07	&4.07\\
$a_x$ (\AA)   				& 3.94  & 4.07	& 4.22	&4.37 & 3.94& 4.07&4.22	&4.37\\
\hline
$\epsilon_{+++}$ 			&  112	& 82	& 87  	&100  & 73  & 52  & 59	&73\\
$\epsilon_{-++}$    			&  200	& 151 	& 139 	&140  & 130 & 95 & 92  &98\\
$\epsilon_{+-+},\epsilon_{++-}$		&  200	& 151	& 156 	&167  & 130 & 95 & 103  &116\\
$\epsilon_{-+-},\epsilon_{--+}$    	& 279	& 211 	& 200	&200  & 183 & 135 & 131  &137\\
$\epsilon_{+--}$ 			&  279	& 211	& 216  	&226  & 183 & 135 & 142  &155\\
$\langle +++| x |-++\rangle$		& 0.15	& 0.17 & 0.20	&0.22&0.13&0.15& 0.18&0.20\\
\hline
$-4.4e^2 d^2 a^{-3}_y \lambda$ 		&  -24  & -28  	& -37   & -47 &  -18& -22 &-29     &-38\\
$\epsilon_{exc}$ 			& 44   	&34.5	& 38    & 48  & 28.5& 21.5&28.5   &40.5\\
\hline
\end{tabular}
\\[1cm]
The state energies are all in meV relative to the lattice potential at the O-site. For  $a_\parallel > a_\perp = 4.07$ \AA, a lattice strain energy contribution, deduced from the Pd-Pd force constant given by \cite{Rah76}, has been added to all states. The penultimate row tabulates the dipole-dipole energy per adatom at maximum alignment and $\lambda=1$. The final row tabulates the corresponding excitation energy per adatom for the pair state $\epsilon_{-}$ in (\ref{epm}) relative to the ground state in the undistorted lattice.
\\
It can be seen from our results for $\lambda=1$ that lattice expansion along the dipole axis results in a rapid increase in dipole-dipole energy and a simultaneous decrease in excitation energy. Moreover, the maximum dipole energy appears to closely match the excitation cost of the oscillations for a wide range of lattice strain values. On the basis of our data, the coherent oscillating phase would appear to be (just) favoured in the deuteride at low temperatures, but in view of the inherent uncertainty in our data, it is not considered worth applying the method of Section 3 to deduce a phase transition temperature.
\section{Conclusion}
We have demonstrated that conventional models of the hydrogen-lattice system have grossly underestimated the effective coupling between neighbouring hydrogen sites. We have furthermore shown that, in one metal at least, the neglected effect may give rise to a new low temperature phase in which all hydrogen adatoms oscillate coherently. The posited effect has the potential to explain both the anomalous shoulder in the optical phonon spectrum and the quantum entanglement indicated by neutron scattering and EELS studies  and we therefore deem that further theoretical investigation is warranted. A further possible consequence might be an enhanced probability of fusion of deuterons in the collective state. Owing to the translational symmetry of the many-body wavefunction across lattice cells, we speculate that the dominant decay channel might be radiationless.

\end{document}